\documentclass[twocolumn,secnumarabic,amssymb,nobibnotes,aps,prl,showpacs,superscriptaddress,longbibliography]{revtex4-2}
\usepackage[english]{babel}
\usepackage{graphicx}
\usepackage{lmodern}
\usepackage{amsmath}
\usepackage{amsfonts}
\usepackage[usenames,dvipsnames]{color}
\usepackage{hyperref}
\hypersetup{colorlinks}
\hypersetup{citecolor=blue}
\usepackage{blindtext}
\usepackage[utf8]{inputenc}
\usepackage{color}
\usepackage{mathtools}
\usepackage[normalem]{ulem}
\usepackage{bm}
\mathtoolsset{showonlyrefs=true}

\def\*#1{\mathbf{#1}}
\def\!#1{\mathbf{\hat#1}}

\begin{document}
\title{Dissipative effects in odd viscous Stokes flow around a single sphere}
\author{Jeffrey C. Everts}
        \email{jeffrey.everts@fuw.edu.pl}
    \affiliation{Institute of Theoretical Physics, Faculty of Physics, University of Warsaw, Pasteura 5, 02-093 Warsaw, Poland}
        \affiliation{Institute of Physical Chemistry, Polish Academy of Sciences, 01-224 Warsaw, Poland}
\author{Bogdan Cichocki}
    \affiliation{Institute of Theoretical Physics, Faculty of Physics, University of Warsaw, Pasteura 5, 02-093 Warsaw, Poland}
\date{\today}

\begin{abstract}
Odd viscosity (OV) is a transport coefficient in, for example, fluids of self-spinning (active) particles or electrons in an external magnetic field. The key feature of OV is that it does not contribute to dissipation in two spatial dimensions. In contrast, we explicitly show that in the three-dimensional case, OV can contribute indirectly to dissipation by modifying the fluid flow. We quantify the dissipation rate of a single spherical particle moving through a fluid with OV via an exact analytical solution of the generalised stationary creeping flow equations. Our results provide a novel way to quantify the effects of OV by measuring the solid-body motion of a single spherical particle. Moreover, we explicitly demonstrate how complex fluids can be designed in terms of their rheological properties by mixing passive particles with self-spinning active particles.
\end{abstract}
\maketitle

\underline{\emph{Introduction}.} -- The response of a liquid to local gradients in the fluid velocity ${\bf v(r)}$ is characterized by the linear constitutive relation $\sigma_{\alpha\beta}^\mathrm{V}({\bf r})=\eta_{\alpha\beta\gamma\lambda}\partial_\lambda v_\gamma({\bf r})$, which describes how a generalised flux (the viscous stress tensor $\boldsymbol{\sigma}^\mathrm{V}$) is related to its conjugate generalised force (the velocity gradient tensor $\nabla{\bf v}$) via a transport coefficient (the viscosity tensor $\boldsymbol{\eta}$). In the presence of external fields ${\bf B}$ that are odd under a time-reversal operation (e.g., a magnetic field), microscopic time reversibility  dictates that the form of $\boldsymbol{\eta}$ is constrained by the Onsager-Casimir reciprocal relation (OCRR) \cite{Onsager:1931a, Onsager:1931b, Casimir:1945}, $\eta_{\alpha\beta\gamma\lambda}({\bf B})=\eta_{\gamma\lambda\alpha\beta}(-{\bf B})$. Consequently, the viscosity tensor picks up a non-vanishing antisymmetry part for ${\bf B}\neq 0$, with $\boldsymbol{\eta}({\bf B})=\boldsymbol{\eta}^\mathrm{S}({\bf B})+\boldsymbol{\eta}^\mathrm{A}({\bf B})$, where $\eta^\mathrm{S}_{\alpha\beta\gamma\lambda}({\bf B})=\eta^\mathrm{S}_{\gamma\lambda\alpha\beta}({\bf B})$ and  $\eta^\mathrm{A}_{\alpha\beta\gamma\lambda}({\bf B})=-\eta^\mathrm{A}_{\gamma\lambda\alpha\beta}({\bf B})$. 

In active matter \cite{Julicher:2018}, there are fluids described by a non-vanishing intrinsic angular momentum density $\boldsymbol{\ell}$, which can be  generated by rotating the system \cite{Delplace:2017}, by using self-spinning particles due to fuel consumption \cite{Glotzer:2014, Maggi:2015, Aubret:2018, Yang:2020, Han:2021}, or external fields \cite{Miloh:2014, Kokot:2017}. So-called chiral active fluids are described by a ground state with  $\boldsymbol{\ell}\neq 0$ that breaks time-reversal symmetry. However, care has to be taken when selecting a \emph{single} ground state and using the OCRR with $\boldsymbol{\ell}$ fixed (although this is common practice). Namely, in the derivation of the OCRR, microscopic time reversibility is assumed on the level of particles, and therefore, the full dynamical properties of $\boldsymbol{\ell}$ have to be included. Viewing $\boldsymbol{\ell}$ as a dynamically-generated order parameter, we have $\boldsymbol{\mathcal{\ell}}\rightarrow -\boldsymbol{\mathcal{\ell}}$ under time reversal and an OCRR in the form $\eta_{\alpha\beta\gamma\lambda}(\boldsymbol{\ell})=\eta_{\gamma\lambda\alpha\beta}(-\boldsymbol{\ell})$. Consequently, there is a non-vanishing $\boldsymbol{\eta}^\mathrm{A}$ similar to a fluid in an external magnetic field. This makes chiral active fluids distinctively different from, for example, cholesteric liquid crystals \cite{Prost1993} (which have $\boldsymbol{\eta}^\mathrm{A}=0$) that are chiral due to their spatial ordering, but not active chiral due to the underlying time-reversal-invariant order parameter.

The coefficients in $\boldsymbol{\eta}^\mathrm{A}$ are known as OV \cite{Avron:1988}, but terms like gyroviscosity \cite{Ramos:2005} and Hall viscosity \cite{Scaffidi:2017} are also used depending on the system of interest, like magnetised plasmas \cite{Stacey:1989} or electrons in graphene \cite{Bandurin:2019}, respectively. The effects of OV can be profound: it can generate edge flows like the edge states in topological insulators \cite{Souslov:2019}, anomalous Hall flows \cite{Lou:2022}, transverse flow patterns to an applied pressure gradient \cite{Banerjee:2017}, or azimuthal flow when a particle is pulled by an external force \cite{Khain:2022}. See Ref. \cite{Fruchart:2023} for a comprehensive literature review. Recently, in soft condensed matter, an OV coefficient is realised with a similar magnitude as the shear viscosity $\eta_\mathrm{s}$ in systems of self-spinning magnetic particles \cite{Soni:2019}, contrasting magnetised polyatomic gases where OV coefficients are orders of magnitude smaller than $\eta_\mathrm{s}$ \cite{Hulsman:1970}. Realisations of OV are also expected to be present in biological matter \cite{Friedrich:2007, Petroff:2015, Fakhri:2022}.

\begin{figure}
\begin{center}
\includegraphics[width=0.45\textwidth]{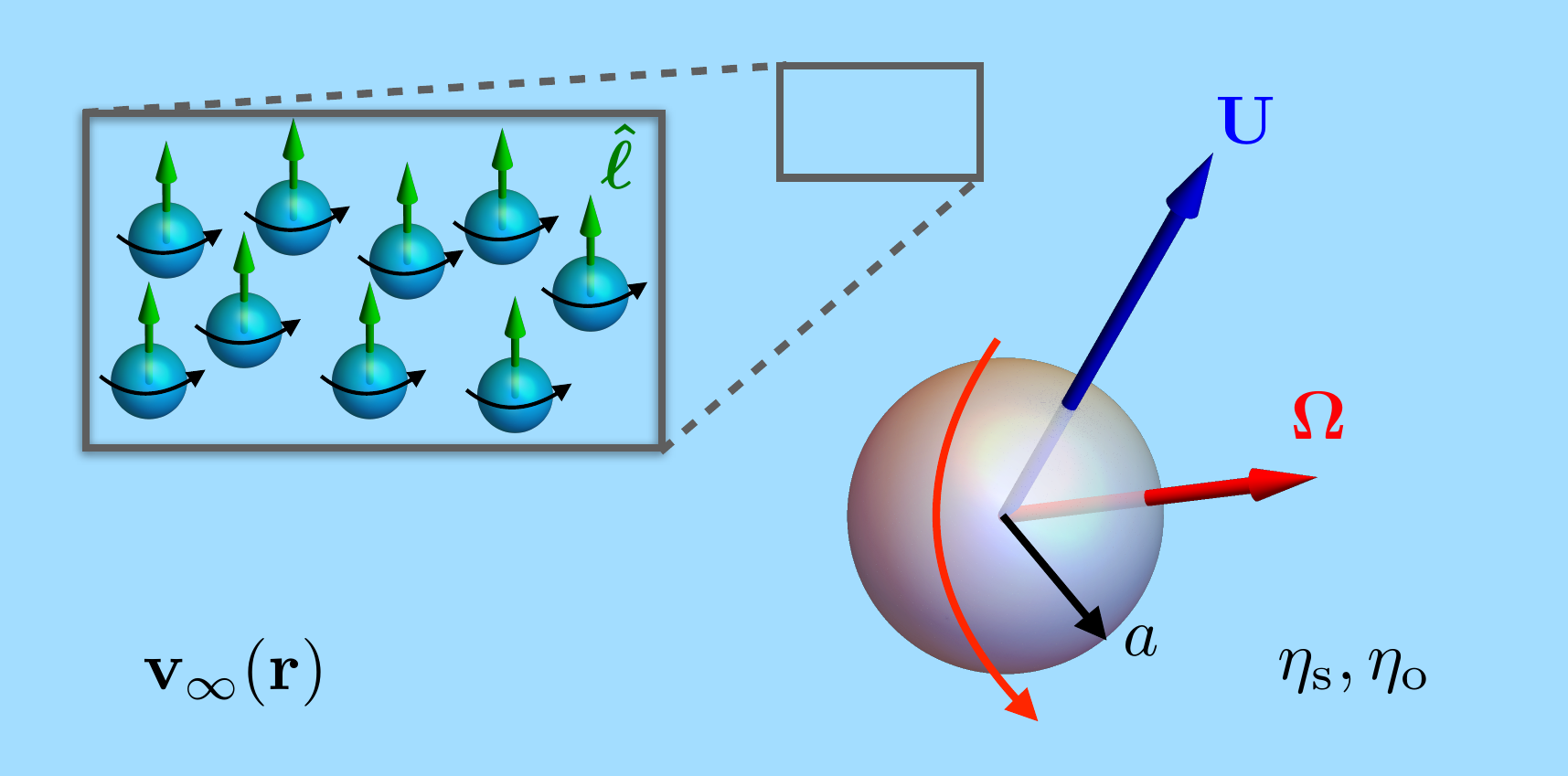} 
\end{center}
\caption{Scheme of a spherical particle of radius $a$ translating with velocity ${\bf U}$ and rotating with angular velocity $\boldsymbol{\Omega}$ in a fluid with shear viscosity $\eta_\mathrm{s}$ and OV $\eta_\mathrm{o}$. The zoomed-in inset shows how OV emerges due to the self-spinning (active) particles in the medium that all rotate in the same direction, denoted by the intrinsic angular momentum direction $\boldsymbol{\hat{\ell}}$. Furthermore, there is a linear ambient flow field ${\bf v}_\infty({\bf r})$ infinitely far from the particle.} \label{fig:scheme}
\end{figure}

Generally, OV does not contribute to the viscous dissipation rate ${\dot{w}({\bf r})}= \sigma_{\alpha\beta}^\mathrm{V}({\bf r})\partial_\beta v_\alpha({\bf r})=\eta_{\alpha\beta\gamma\lambda}^\mathrm{S}\partial_\beta v_\alpha({\bf r}) \partial_\lambda v_\gamma({\bf r})$, provided that ${\bf v(r)}$ is not affected by OV. Notably, in two spatial dimensions, the effects of OV can be absorbed in the pressure for incompressible flow and hence do not contribute to the flow velocity \cite{Abanov:2017} (contrasting compressible fluids \cite{Lier:2023} and other systems coupled to a bulk passive fluid \cite{Hosaka:2021}). However, in three spatial dimensions, OV can affect fluid flow and hence indirectly contribute to viscous dissipation \cite{Khain:2022}. There is a lacking quantification of this effect because the simplest three-dimensional system where this effect is present -- a single sphere undergoing solid-body motion in an OV liquid in the stationary creeping flow regime-- has only been investigated in the limit of small OV (with respect to $\eta_\mathrm{s}$) where such dissipative effects identically vanish \cite{Khain:2022, Cruz:2023}. This limits the experimental applicability to, for example, the system in Ref. \cite{Soni:2019}, where $\eta_\mathrm{o}\approx\eta_\mathrm{s}$.

In this Letter, we quantify the dissipation rate of a single spherical particle undergoing solid-body motion in an incompressible active chiral fluid  in linear ambient flow, see Fig. \ref{fig:scheme}. We demonstrate that for this situation, an exact analytical solution exists for the relation between force moments and the solid-body motion of a single sphere. Furthermore, we discuss the dissipated power due to OV and its effects on effective hydrodynamic equations for suspensions of spherical particles {\color{black} in chiral active fluids, a concept not yet considered before within this context in the literature}. Going beyond the regime of weak OV, we lay the building blocks for describing Brownian motion, hydrodynamic interactions, and effective hydrodynamic equations of colloidal suspensions in a three-dimensional chiral active fluid.

\underline{\emph{Model}.} -- Consider an incompressible liquid, $\nabla\cdot{\bf v}({\bf r})=0$ described by the stress tensor $\boldsymbol{\sigma}({\bf r})=-p({\bf r})\bm{\mathsf{I}}+\boldsymbol{\eta}:\nabla{\bf v}({\bf r})$, where we defined the single contraction as ${[\bm{\mathsf{A}}\cdot\bm{\mathsf{B}}]_{\alpha...\beta}={A}_{\alpha...\lambda}{B}_{\lambda...\beta}}$ and the double contraction as $[\bm{\mathsf{A}}:\bm{\mathsf{B}}]_{\alpha...\beta}={A}_{\alpha...\gamma\lambda}{B}_{\lambda\gamma...\beta}$. Furthermore, $[\bm{\mathsf{I}}]_{\alpha\beta}=\delta_{\alpha\beta}$. In the absence of an external magnetic field, we choose the simplest form of the viscosity tensor $\boldsymbol{\eta}$ (for the general form see Ref. \cite{Khain:2022}) assuming {\color{black} that there is only one type of shear viscosity $\eta_\mathrm{s}$}, but allowing for a constant non-zero {\color{black} intrinsic} angular momentum density $\boldsymbol{\ell}$,
\begin{equation}
\boldsymbol{\eta}=2\eta_\mathrm{s}\boldsymbol{\mathcal{I}}+{\color{black}2\eta_\mathrm{o}\bm{\mathsf{c}}^{(0)}(\boldsymbol{\hat{\ell}})}, \label{eq:visctens}
\end{equation}
Eq. \eqref{eq:visctens} has been derived from a microscopic Hamiltonian of self-spinning particles in Ref. \cite{Markovich:2021} {\color{black} and is experimentally relevant \cite{Soni:2019}}. Here, we defined the tensors $\mathcal{I}_{\alpha\beta\sigma\nu}=\delta_{\alpha(\sigma}\delta_{\nu)\beta}-(1/3)\delta_{\alpha\beta}\delta_{\sigma\nu}$ and {\color{black} $c_{\alpha\beta\sigma\nu}^{(0)}=\hat{\ell}_\lambda(\epsilon_{\lambda\alpha(\sigma}\delta_{\nu)\beta}+\epsilon_{\lambda\beta(\sigma}\delta_{\nu)\alpha})$}, with indices between round brackets indicating the symmetric part. For example, $\delta_{\alpha(\sigma}\delta_{\nu)\beta}=(\delta_{\alpha\sigma}\delta_{\nu\beta}+\delta_{\alpha\nu}\delta_{\sigma\beta})/2$. In Eq. \eqref{eq:visctens}, we characterize the OV by the length of $\boldsymbol{\ell}$ via $\eta_\mathrm{o}>0$, and its direction $\boldsymbol{\hat{\ell}}$. This is equivalent to the frequently used convention of fixing $\boldsymbol{\hat{\ell}}$ but allowing $\eta_\mathrm{o}$ to be negative. In our formulation, however, the OCRR are manifest by including the dynamical character of $\boldsymbol{\ell}$.

Furthermore, in Eq. \eqref{eq:visctens}, we assume that there are no torque densities; hence, $\boldsymbol{\sigma}$ is symmetric. Antisymmetric stresses can be, in principle, included \cite{Felderhof:2011}, {\color{black} as well as having an inhomogeneous $\boldsymbol{\ell}$ \cite{Markovich:2021}}. For stationary creeping flow, conservation of linear momentum $\nabla\cdot\boldsymbol{\sigma}({\bf r})=-{\bf f (r)}$, with ${\bf f(r)}$ a force density acting on the fluid, gives
\begin{gather}
\eta_\mathrm{s}\nabla^2{\bf v({\bf r}})-\nabla\tilde{p}({\bf r})+\eta_\mathrm{o}(\boldsymbol{\hat{\ell}}\cdot\nabla)[\nabla\times{\bf v}({\bf r})]=-{\bf f}({\bf r}). \label{eq:gensto} 
\end{gather}
 {\color{black} In Eq. \eqref{eq:gensto}, the intrinsic angular momentum of the fluid $\boldsymbol{\hat{\ell}}$ couples to the orbital angular momentum of the fluid $\nabla\times{\bf v}({\bf r})$.}
We extracted a total gradient from the viscous stress tensor to define the effective pressure, $\tilde{p}({\bf r})=p({\bf r})+2\eta_\mathrm{o}\boldsymbol{\hat{\ell}}\cdot[\nabla\times{\bf v}({\bf r})]$. This form is chosen because the incompressibility condition leads to $\nabla^2\tilde{p}({\bf r})=\nabla\cdot{\bf f}({\bf r})$, which is of the Stokes form.

\underline{\emph{Fundamental solution}.} -- First, we construct the fundamental solution to Eq. \eqref{eq:gensto}, which is defined as the response to a point force density. The corresponding Green tensor $\bm{\mathsf{G}}({\bf r})$ and effective pressure vector $\bar{{\bf Q}}({\bf r})$ satisfy
\begin{gather}
\eta_\mathrm{s}\nabla^2 G_{\alpha\beta}({\bf r})-\partial_{\alpha}\bar Q_\beta({\bf r})+\eta_\mathrm{o}\epsilon_{\alpha\gamma\lambda}(\boldsymbol{\hat{\ell}}\cdot\nabla)\partial_\gamma G_{\lambda\beta}({\bf r})\nonumber \\=-\delta_{\alpha\beta}\delta({\bf r}), \quad \partial_\alpha G_{\alpha\beta}({\bf r})=0. \label{eq:green}
\end{gather}
Insertion of $\bm{\mathsf{G}}({\bf r})=\int d{\bf k}/(2\pi)^3\, \tilde{\bm{\mathsf{G}}}({\bf k})e^{i{\bf k}\cdot{\bf r}}$, with $\tilde{\bm{\mathsf{G}}}({\bf k})$ the Fourier transform of $\bm{\mathsf{G}}({\bf r})$ (and with similar definitions for $\tilde{{\bar{\bf {Q}}}}({\bf k})$), we find upon using the incompressibility condition that $\tilde{\bar{{\bf {Q}}}}({\bf k})=-i\hat{\bf k}/k$. Insertion back in Eq. \eqref{eq:green} gives 
\begin{equation}
\tilde{\bm{\mathsf{G}}}({\bf k})=\frac{1}{\eta_\mathrm{s}}\frac{\bm{\mathsf{I}}-\hat{\bf k}\hat{\bf k}+\gamma (\hat{\bf k}\cdot\boldsymbol{\hat{\ell}})(\boldsymbol{\epsilon}\cdot\hat{\bf k})}{k^2[1+\gamma^2(\hat{\bf k}\cdot\boldsymbol{\hat{\ell}})^2]}, \label{eq:ft}
\end{equation}
with $\boldsymbol{\epsilon}$ the Levi-Cevita symbol and $\gamma=\eta_\mathrm{o}/\eta_\mathrm{s}$. In real space, we conclude that $\bar{\bm{\mathsf{Q}}}({\bf r})=\hat{\bf r}/(4\pi r^2)$, (like in the Stokes case). The Fourier inverse of Eq. \eqref{eq:ft} has been calculated to linear order in $\gamma$ in Ref. \cite{Cruz:2023} and only for components $G_{\alpha z}$ ($\alpha=x,y,z$) in Ref. \cite{Khain:2022}. The full tensor can be calculated by using cylindrical coordinates where $\hat{\boldsymbol{\ell}}$ is the longitudinal axis, followed by contour integration over the axial coordinate \cite{Khain:2022}. The remaining integrals over the tangential and radial coordinate are known \cite{Gradsteyn}; see SM \cite{SM} for technical details. We find
\begin{gather}
\bm{\mathsf{G}}({\bf r})=\frac{1}{4\pi\eta_\mathrm{s}(r+\tilde{r})}\Bigg\{\bm{\mathsf{I}}+\hat{\bf r}\hat{\bf r}+\frac{\gamma r}{\tilde{r}}[(\hat{\boldsymbol{\ell}}\cdot\boldsymbol{\epsilon})-(\hat{\bf r}\cdot\hat{\boldsymbol{\ell}})(\hat{\bf r}\cdot\boldsymbol{\epsilon})]
\nonumber \\
-\left(1-\frac{r}{\tilde{r}}\right)\left[\hat{\bf r}\hat{\bf r}+\frac{(\hat{\bf r}\times\hat{\boldsymbol{\ell}})(\hat{\bf r}\times\hat{\boldsymbol{\ell}})}{1-(\hat{\bf r}\cdot\hat{\boldsymbol{\ell}})^2}\right]\Bigg\}, \label{eq:Greenreal}
\end{gather}
with $\tilde{r}/r=\sqrt{1+\gamma^2[1-(\hat{\bf r}\cdot\hat{\boldsymbol{\ell}})^2]}$. Note that we have the symmetry relation $G_{\alpha\beta}({\bf r};\boldsymbol{\hat{\ell}})=G_{\beta\alpha}({\bf r};-\boldsymbol{\hat{\ell}})$ and for $\gamma\rightarrow 0$, we have $\tilde{r}\rightarrow r$ and, therefore, we retrieve the Oseen tensor: $\lim_{\gamma\rightarrow 0}\bm{\mathsf{G}}({\bf r})=(\bm{\mathsf{I}}+\hat{\bf r}\hat{\bf r})/(8\pi\eta_\mathrm{s}r)$.

\underline{\emph{Problem formulation}.} -- Consider a sphere of radius $a$ in our OV model undergoing solid-body motion with stick boundary conditions
\begin{equation}
{\bf v}(a\hat{\bf r})={\bf U}+\boldsymbol{\Omega}\times(a\hat{\bf r}), \label{eq:bc}
\end{equation}
where ${\bf U}$ is a constant translational velocity and $\boldsymbol{\Omega}$ is the rotational velocity, see Fig. \ref{fig:scheme}. Furthermore, in this coordinate system, the origin coincides with the centre of the sphere. We include the presence of an ambient flow field ${\bf v}^\infty({\bf r})={\bf U}^\infty+\boldsymbol{\Omega}^\infty\times{\bf r}+\bm{\mathsf{E}}^\infty\cdot{\bf r}$, containing a constant ambient flow ${\bf U}^\infty$ describing translation, and linear ambient flows (i.e., $\sim{\bf r}$), with $\boldsymbol{\Omega}^\infty$ denoting constant rotation and the symmetric traceless tensor $\bm{\mathsf{E}}^\infty$ describing linear shear flow, of the fluid at infinity, i.e., $\lim_{r\rightarrow\infty}[{\bf v(r)}-{\bf v}^\infty({\bf r})]=0$. Our goal is to solve Eq. \eqref{eq:gensto} (with ${\bf f}=0$) with Eq. \eqref{eq:bc} as the boundary condition. Inserting the resulting ${\bf v}({\bf r})$ in $\boldsymbol{\sigma}({\bf r})$ allows us to compute the force ${\bf F}$, torque ${\bf T}$ and stresslet $\bm{\mathsf{S}}$ exerted by the fluid on the spherical particle \cite{Kim}, 
\begin{gather}
{\bf F}=\oint_{r=a}dS\, \boldsymbol{\sigma}({\bf r})\cdot\hat{\bf r}, \quad
{\bf T}=\oint_{r=a}dS\, {\bf r}\times[\boldsymbol{\sigma}({\bf r})\cdot\hat{\bf r}], \label{eq:fmoments}\\
\bm{\mathsf{S}}=\oint_{r=a}dS\,\frac{1}{2}\left\{ {\bf r}[\boldsymbol{\sigma}({\bf r})\cdot\hat{\bf r}]+[\boldsymbol{\sigma}({\bf r})\cdot\hat{\bf r}]{\bf r}-\frac{2}{3}{\bf r}\cdot[\boldsymbol{\sigma}({\bf r})\cdot\hat{\bf r}]\bm{\mathsf{I}}\right\}, \nonumber
\end{gather}
with $\hat{\bf r}$ the outward pointing unit normal for a sphere. 

In order to describe diffusion and effective viscosities, we need to compute the grand mobility matrix $\boldsymbol{\mu}$, defined by the linear relation
\begin{gather}
\begin{pmatrix}
{\bf U}-{\bf U}^\infty \\
\boldsymbol{\Omega}-\boldsymbol{\Omega}^\infty\\
\bm{\mathsf{S}}
\end{pmatrix}
=-
\begin{pmatrix}
\boldsymbol{\mu}^\mathrm{tt} & \boldsymbol{\mu}^\mathrm{tr} & \boldsymbol{\mu}^\mathrm{td} \\
\boldsymbol{\mu}^\mathrm{rt} & \boldsymbol{\mu}^\mathrm{rr} & \boldsymbol{\mu}^\mathrm{rd} \\
\boldsymbol{\mu}^\mathrm{dt} & \boldsymbol{\mu}^\mathrm{dr} & \boldsymbol{\mu}^\mathrm{dd}
\end{pmatrix}
\begin{pmatrix}
{\bf F} \\
{\bf T}\\
-\bm{\mathsf{E}}^\infty
\end{pmatrix}. \label{eq:muuu}
\end{gather}
Here, the matrix product needs to be interpreted accordingly. 
For example, the last line is $\bm{\mathsf{S}}=-(\boldsymbol{\mu}^\mathrm{dt}\cdot{\bf F}+\boldsymbol{\mu}^\mathrm{dr}\cdot{\bf T}-\boldsymbol{\mu}^\mathrm{dd}:\bm{\mathsf{E}}^\infty)$. For the solution of the friction problem defined by Eqs. \eqref{eq:gensto} and \eqref{eq:bc} and for the calculation of the viscous dissipation rate, the grand resistance matrix $\boldsymbol{\zeta}$ is relevant, which is defined as the partial inversion of $\boldsymbol{\mu}$,
\begin{gather}
\begin{pmatrix}
{\bf F} \\
{\bf T}\\
{\bm{\mathsf{S}}}
\end{pmatrix}
=-
\begin{pmatrix}
\boldsymbol{\zeta}^\mathrm{tt} & \boldsymbol{\zeta}^\mathrm{tr} & \boldsymbol{\zeta}^\mathrm{td} \\
\boldsymbol{\zeta}^\mathrm{rt} & \boldsymbol{\zeta}^\mathrm{rr} & \boldsymbol{\zeta}^\mathrm{rd} \\
\boldsymbol{\zeta}^\mathrm{dt} & \boldsymbol{\zeta}^\mathrm{dr} & \boldsymbol{\zeta}^\mathrm{dd}
\end{pmatrix}
\begin{pmatrix}
{\bf U}-{\bf U}^\infty \\
\boldsymbol{\Omega}-\boldsymbol{\Omega}^\infty\\
-{\bm{\mathsf{E}}}^\infty
\end{pmatrix}. \label{eq:z}
\end{gather}
As we show below, there are no translational-rotational and translational-dipolar couplings. 

\underline{\emph{Translating sphere in constant ambient flow}.} -- 
With Eq. \eqref{eq:Greenreal}, we use the singularity method proposed by Brenner \cite{Brenner:1964, Brenner:1964b, Brenner:1966} to find a solution to the friction problem defined by Eqs. \eqref{eq:gensto} and \eqref{eq:bc} for $\boldsymbol{\Omega}=0$ and ${\bf v}^\infty({\bf r})={\bf U}^\infty$. We denote the corresponding velocity field by ${\bf v}_0({\bf r})$. Following his considerations, we find that
\begin{equation}
{\bf v}_0({\bf r})-{\bf U}^\infty=[\mathcal{L}_0\bm{\mathsf{G}}]({\bf r})\cdot\boldsymbol{\zeta}^\mathrm{tt}\cdot({\bf U}-{\bf U}^\infty), \label{eq:solU}
\end{equation}
with a linear operator $\mathcal{L}_0$ given by
\begin{equation}
\mathcal{L}_0=\sum_{n=0}^\infty\frac{a^{2n}}{(2n+1)!}(\nabla^2)^n=:j_0(i\mathcal{D}). \label{eq:L0}
\end{equation}
Here, the righthandside formally defines a series expansion in terms of $\mathcal{D}$, where $\mathcal{D}^2=a^2\nabla^2$ and $j_n$ is $n$-th order spherical Bessel function of the first kind. 
Clearly, Eq. \eqref{eq:solU} solves Eq. \eqref{eq:gensto}. To avoid repeating the tedious steps from Refs. \cite{Brenner:1964, Brenner:1964b, Brenner:1966}, we only need to show that the boundary condition Eq. \eqref{eq:bc} is satisfied, which determines the yet unknown $\boldsymbol{\zeta}^\mathrm{tt}$. 

We find $\bm{\mathsf{G}}({\bf r})$ in terms of its Fourier transform, \begin{equation}
[\mathcal{L}_0\bm{\mathsf{G}}]({\bf r})=\int\frac{d{\bf k}}{(2\pi)^3}\, e^{i{\bf k}\cdot{\bf r}}j_0(ka)\bm{\mathsf{G}}({\bf k}). \label{eq:vU}
\end{equation}
Observe that $j_0(ka)$ does not contain any poles on the axial coordinate in appropriate cylindrical coordinates. Therefore, we can use the same method as how we determined Eq. \eqref{eq:Greenreal} from Eq. \eqref{eq:ft} to evaluate Eq. \eqref{eq:vU}. We find that $[\mathcal{L}_0\bm{\mathsf{G}}](a\hat{\bf r})$ is independent of the direction $\hat{\bf r}$, which means that we can satisfy the boundary condition ${\bf v}_0(a\hat{\bf r})={\bf U}$ if $\boldsymbol{\zeta}^\mathrm{tt}=[\boldsymbol{\mu}^\mathrm{tt}]^{-1}$, with the identification $\boldsymbol{\mu}^\mathrm{tt}=[\mathcal{L}_0\bm{\mathsf{G}}](a\hat{\bf r})$. We find by evaluating Eq. \eqref{eq:vU} for ${\bf r}=a\hat{\bf r}$ (see SM \cite{SM})
\begin{gather}
\boldsymbol{\mu}^\mathrm{tt}=\frac{1}{24\pi\eta_\mathrm{s}a}\Big\{[\gamma^2n(\gamma)+4](\bm{\mathsf{I}}-\boldsymbol{\hat{\ell}\hat{\ell}})\nonumber\\
+[2\gamma^2m(\gamma)+4]\boldsymbol{\hat{\ell}\hat{\ell}}-2\gamma f(\gamma)(\boldsymbol{\epsilon}\cdot\boldsymbol{\hat{\ell}})\Big\}, \label{eq:mutt}
\end{gather}
with $m(\gamma)=f(\gamma)+g(\gamma)$, $n(\gamma)=f(\gamma)-g(\gamma)$. Here, $f(\gamma)=(3/\gamma^2)(\psi/\gamma-1)=-1+\mathcal{O}(\gamma^2)$ and ${g(\gamma)=[1+f(\gamma)]/\gamma^2=3/5+\mathcal{O}(\gamma^2)}$.  Furthermore, we defined the ``angle"  $\psi=\mathrm{arcsin}(\gamma/\sqrt{1+\gamma^2})$. Note that $\mu^\mathrm{tt}_{\alpha\beta}(\boldsymbol{\hat{\ell}})=\mu^\mathrm{tt}_{\beta\alpha}(-\boldsymbol{\hat{\ell}})$ is a consequence of the OCRR.
 Eq. \eqref{eq:mutt} implies that pulling a sphere with a constant force will generate components of ${\bf U}$ that are perpendicular to the direction of the force,  in contrast to the case $\gamma=0$, where ${\bf U}\parallel {\bf F}$. 

Similarly, the method to obtain explicit expressions for ${\bf v}_0({\bf r})$ by using Eq. \eqref{eq:vU} is given in the SM \cite{SM}. Here, we just quote the result for ${\bf U}=U\hat{\boldsymbol{\ell}}$ and ${\bf U}^\infty=0$,
\begin{align}
{\bf v}_0({\bf r})&=\frac{3aU}{\gamma^2m(\gamma)+2}\Bigg\{\left[\frac{\mathrm{arcsin}\left(\mathcal{R}_+^{-1}\right)}{\gamma a\sin^2\psi}-\frac{1}{\gamma^2 r}\right]\boldsymbol{\hat{\ell}} \label{eq:cool}\\
&+\frac{{\hat{\bf r}}\cdot\hat{\boldsymbol{\ell}}}{\gamma^2r\sqrt{1-(\hat{\bf r}\cdot\boldsymbol{\hat\ell})^2}}\left(\frac{\sqrt{1-\mathcal{R}_-^2}}{|{\hat{\bf r}}\cdot\hat{\boldsymbol{\ell}}|}-1\right)(\gamma\boldsymbol{\hat\phi}-\boldsymbol{\hat{\rho}})\Bigg\}, 
\end{align}
where $\mathcal{R}_\pm=(\mathcal{A}_+\pm\mathcal{A}_-)/(2a\sin\psi)$, with 
\begin{equation}
\mathcal{A}_\pm=\sqrt{({\bf r}\cdot\boldsymbol{\hat\ell})^2\cos^2\psi +\left[r\sqrt{1-(\hat{\bf r}\cdot\boldsymbol{\hat\ell})^2}\pm a\sin\psi\right]^2},
\end{equation}
with corresponding effective pressure ${\tilde{p}_0({\bf r})=\tilde{p}^\infty+\bar{\bm{\mathsf{Q}}}({\bf r})\cdot\boldsymbol{\zeta}^\mathrm{tt}\cdot({\bf U}-{\bf U}^\infty)}$, for some constant $\tilde p^\infty$.
Eq. \eqref{eq:cool} coincides with the Stokes result for $\gamma\rightarrow 0$ and to linear order in $\gamma$ with the result of Ref. \cite{Khain:2022}. Moreover, OV generates an axial flow field that is absent for $\gamma=0$. Furthermore, we obtain the stress tensor corresponding to the singularity representation of the velocity field Eq. \eqref{eq:solU}. Upon insertion in Eq. \eqref{eq:fmoments} and the straightforward use of integral theorems \cite{Kim}, we find that ${\bf T}=0$ and ${\bm{\mathsf{S}}}=0$ (see SM for details): we conclude that all tr, td, rt, and dt elements vanish in Eqs. \eqref{eq:muuu} and  \eqref{eq:z}.

\begin{figure}[t]
\begin{center}
\includegraphics[width=0.47\textwidth]{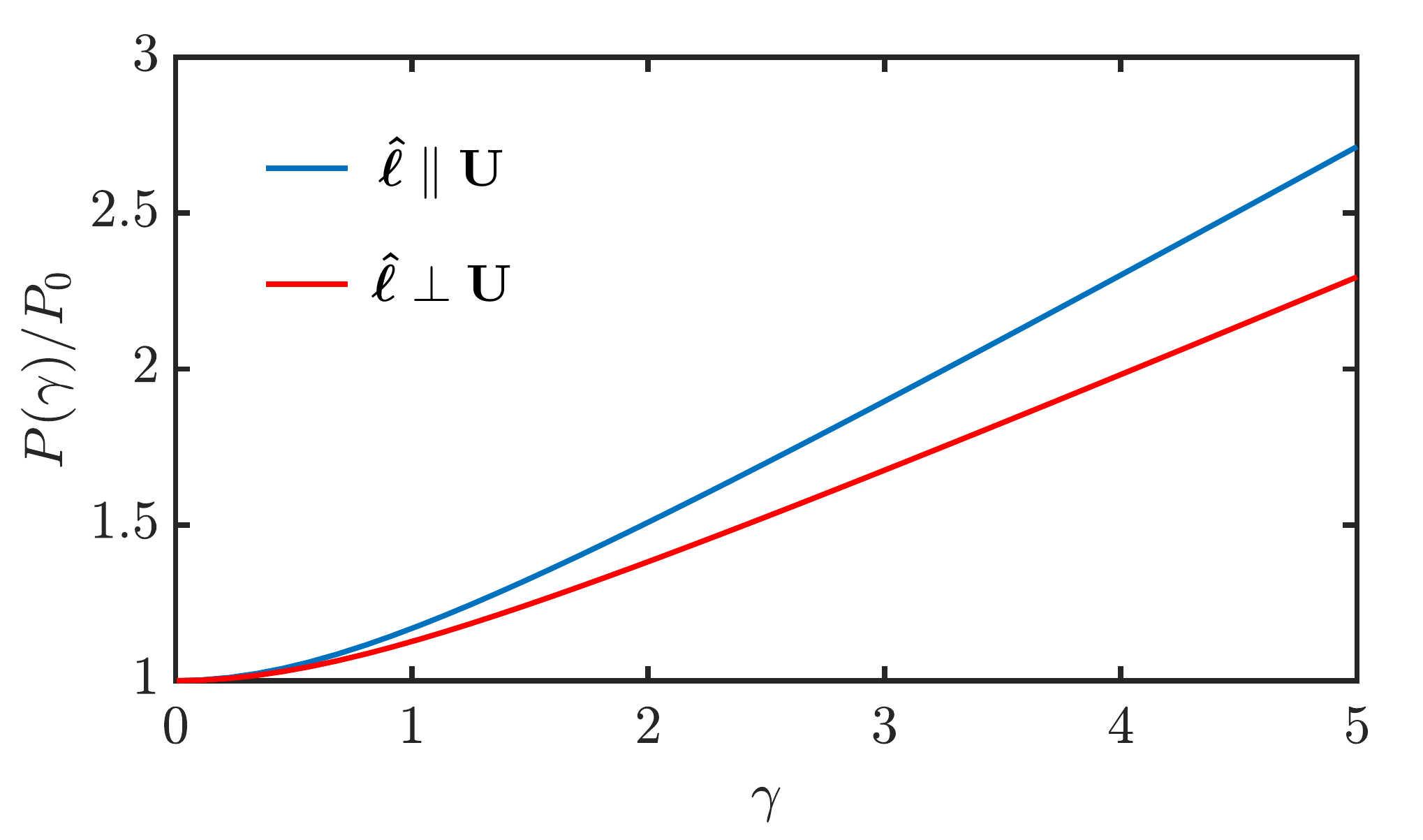} 
\end{center}
\caption{Dissipation rate $P_\mathrm{T}(\gamma)$ with respect to $P_\mathrm{T,0}:=P_\mathrm{T}(\gamma=0)$ of a sphere translating with constant velocity ${\bf U}$ in a chiral active fluid characterised by the ratio $\gamma=\eta_\mathrm{o}/\eta_\mathrm{s}$. For a translating sphere, the dissipation is largest when $\boldsymbol{\hat{\ell}}\parallel{\bf U}$ (blue line) and smallest when $\boldsymbol{\hat{\ell}}\perp{\bf U}$ (red line). For a rotating sphere there is no additional dissipation compared to the ordinary Stokes dissipation, see main text. } \label{fig:dissipation}
\end{figure}

Next, we compute the dissipation rate $P_\mathrm{T}(\gamma)=\int_{r>a}d{\bf r}\, \dot{w}({\bf r})$ for given ${\bf U}$. Here, $P_T(\gamma)$ can be expressed in terms of the work done on the particle surface \cite{Hill:1956, Kim}, and we find $P_\mathrm{T}(\gamma)=-{\bf U}\cdot{\bf F}=\zeta_{(\alpha\beta)}^\mathrm{tt}U_\alpha U_\beta. 
$ There is no direct contribution from the antisymmetric part of $\boldsymbol{\zeta}^\mathrm{tt}$ to $P_\mathrm{T}(\gamma)$. However, there is an indirect contribution of OV to $P_\mathrm{T}(\gamma)$ via the fluid velocity field parametrised by the symmetric part of $\boldsymbol{\zeta}^\mathrm{tt}$. In Fig. \ref{fig:dissipation}(a), we plot $P_\mathrm{T}(\gamma)$ with respect to the Stokes dissipation rate $P_{\mathrm{T},0}:={P_\mathrm{T}(\gamma=0)=6\pi\eta_\mathrm{s}a|{\bf U}|^2}$ and we see that the presence of odd viscosity enhances dissipation, which is largest when $\bf U\parallel\boldsymbol{\hat{\ell}}$. Therefore, translational self-diffusion in an OV medium is reduced compared to the Stokes case. 

\underline{\emph{Rotating sphere in linear ambient flow}.} -- To solve the boundary value problem of a rotating sphere (${{\bf U}={\bf U}^\infty=0}$, $\boldsymbol{\Omega}\neq 0$) with velocity field ${\bf v}_1({\bf r})$, we need to include the linear ambient flow because of the rotational-dipolar coupling.
We define the disturbance velocity ${\bf v}_1^\mathrm{D}={\bf v}_1({\bf r})-\boldsymbol{\Omega}^\infty\times{\bf r}-\bm{\mathsf{E}}^\infty\cdot{\bf r}$, and
using again Brenner's considerations \cite{Brenner:1964, Brenner:1964b, Brenner:1966}, we find that
\begin{align}
v_{1,\alpha}^\mathrm{D}({\bf r})=[\mathcal{L}_1\partial_{\nu} G_{\alpha\beta}]({\bf r})\left(\frac{1}{2}\epsilon_{\nu\beta\lambda}T_\lambda
+{S}_{\nu\beta}\right), \label{eq:velovelo}
\end{align}
with
\begin{equation}
\mathcal{L}_1=\frac{3}{i\mathcal{D}}j_1(i\mathcal{D}):=\sum_{n=0}^\infty\frac{6(n+1)}{(2n+3)!}a^{2n}(\nabla^2)^n,
\end{equation}
where ${\bf T}$ and $\bm{\mathsf{S}}$ are related to $\boldsymbol{\Omega-\Omega}^\infty$ and $\bm{\mathsf{E}}^\infty$ via the rotational-dipolar part of $\boldsymbol{\zeta}$, given by Eq. \eqref{eq:z}. The relevant elements of $\boldsymbol{\zeta}$ are to be determined by the boundary conditions.
Using the same steps as for the translating sphere, we find
\begin{equation}
[\mathcal{L}_1\nabla\bm{\mathsf{G}}]({\bf r})=\frac{3}{a}\int\frac{d{\bf k}}{(2\pi)^3}\, j_1(ka) e^{i{\bf k}\cdot{\bf r}}\, \hat{\bf k} \bm{\mathsf{G}}({\bf k}).
\end{equation}
Explicit calculation  reveals that $[\mathcal{L}_1\nabla \bm{\mathsf{G}}](a\hat{\bf r})\sim\hat{\bf r}$, {\color{black} meaning that the singularity representation Eq. \eqref{eq:velovelo} satisfies the boundary condition ${\bf v}_1(a\hat{\bf r})=\boldsymbol{\Omega}\times(a\hat{\bf r})$}.  By linearity of Eq. \eqref{eq:gensto}, we conclude that the full solution of a sphere undergoing constant solid-body motion in an affine ambient flow field is
${\bf v}({\bf r})-{\bf v}^\infty({\bf r})={\bf v}_0({\bf r})+{\bf v}_1({\bf r})$. Furthermore, $[\mathcal{L}_1\nabla \bm{\mathsf{G}}](a\hat{\bf r})\sim\hat{\bf r}$ can be related to elements of $\boldsymbol{\mu}$ in the rotational-dipolar sector. {\color{black} However, the construction is less straightforward than finding $\boldsymbol{\mu}^\mathrm{tt}$, due to the rotational-dipolar coupling: the details are found in the SM \cite{SM}.} The results are
\begin{gather}
\boldsymbol{\mu}^\mathrm{rr}=\frac{1}{8\pi\eta_\mathrm{s}a^3}\left\{\boldsymbol{\hat{\ell}}\boldsymbol{\hat{\ell}}+\left(1+\frac{\gamma^2}{4}\right)^{-1}\left[(\bm{\mathsf{I}}-\boldsymbol{\hat{\ell}\hat{\ell}})+\frac{\gamma}{2}(\boldsymbol{\epsilon}\cdot\boldsymbol{\hat{\ell}})\right]
\right\}, \\
\boldsymbol{\mu}^\mathrm{dr}=-\frac{\gamma}{2}\bm{\mathsf{p}}^{(0)}-\frac{1}{4+\gamma^2}\left(\gamma\bm{\mathsf{p}}^{(1)}+\frac{\gamma^2}{2}\bm{\mathsf{q}}\right), \\
\boldsymbol{\mu}^\mathrm{dd}=-16\pi\eta_\mathrm{s}a^3\Bigg(\frac{\bm{\mathsf{d}}^{(0)}}{6m(\gamma)}+Y(\gamma)\Big\{2\gamma[g(\gamma)+m(\gamma)]\bm{\mathsf{c}}^{(1)}\\
-[4g(\gamma)+\gamma^2k(\gamma)]\bm{\mathsf{d}}^{(1)}\Big\}+Z(\gamma)\Big\{\gamma m(\gamma)\bm{\mathsf{c}}^{(2)}\\
-[\gamma^2f(\gamma)+3-g(\gamma)]\bm{\mathsf{d}}^{(2)}\Big\}\Bigg), \label{eq:mudd}
\end{gather}
with $k(\gamma)=f(\gamma)+3g(\gamma)$ and given functions
\begin{gather}
Y(\gamma)=\left\{4\gamma^2[g(\gamma)+m(\gamma)]^2+[4g(\gamma)+\gamma^2k(\gamma)]^2\right\}^{-1}, \\
Z(\gamma)=\left\{[\gamma^2f(\gamma)+3-g(\gamma)]^2+4\gamma^2m(\gamma)^2\right\}^{-1}.
\end{gather}
Moreover, we introduced the rank-3 axisymmetric basis (pseudo)tensors \cite{Kim}
\begin{gather}
p_{\alpha\beta\nu}^{(0)}=\left(\hat{\ell}_{\alpha}\hat{\ell}_\beta-\frac{1}{3}\delta_{\alpha\beta}\right)\hat{\ell}_\nu, \quad
 p_{\alpha\beta\nu}^{(1)}=2\hat{\ell}_{(\alpha}\delta_{\beta)\nu}-2\hat{\ell}_\alpha\hat{\ell}_\beta\hat{\ell}_\nu,  \nonumber\\ q_{\alpha\beta\nu}=(\epsilon_{\alpha\nu \sigma}\hat{\ell}_\beta+\epsilon_{\beta\nu \sigma}\hat{\ell}_\alpha)\hat{\ell}_\sigma,
\end{gather}
and rank-4 axisymmetric basis tensors \cite{Kim} constrained by $\bm{\mathsf{d}}^{(0)}+\bm{\mathsf{d}}^{(1)}+\bm{\mathsf{d}}^{(2)}=\boldsymbol{\mathcal{I}}$, where
\begin{gather}
d_{\alpha\beta\sigma\nu}^{(0)}=\frac{3}{2}\left(\hat{\ell}_\alpha\hat{\ell}_\beta-\frac{1}{3}\delta_{\alpha\beta}\right)\left(\hat{\ell}_\sigma\hat{\ell}_\nu-\frac{1}{3}\delta_{\sigma\nu}\right), \nonumber \\
d_{\alpha\beta\sigma\nu}^{(1)}=2\left(\hat{\ell}_{(\alpha}\delta_{\beta)(\sigma}\hat{\ell}_{\nu)}-\hat{\ell}_\alpha\hat{\ell}_\beta\hat{\ell}_\sigma\hat{\ell}_\nu\right), \nonumber 
\end{gather}
and pseudotensors \cite{Gaite:2003}  $\bm{\mathsf{c}}^{(2)}=\bm{\mathsf{c}}^{(1)}-\bm{\mathsf{c}}^{(0)}$ with
$
c_{\alpha\beta\sigma\nu}^{(1)}=\hat{\ell}_\lambda(\epsilon_{\lambda\alpha(\sigma}\hat{\ell}_{\nu)}\hat{\ell}_\beta+\epsilon_{\lambda\beta(\sigma}\hat{\ell}_{\nu)}\hat{\ell}_\alpha), 
$ 
and $\bm{\mathsf{c}}^{(0)}$ defined under Eq. \eqref{eq:visctens}.
We have the symmetry relations $\mu^\mathrm{rr}_{\alpha\beta}(\boldsymbol{\hat{\ell}})=\mu^\mathrm{rr}_{\beta\alpha}(-\boldsymbol{\hat{\ell}})$, ${\mu}_{\alpha\beta\nu}^\mathrm{rd}(\boldsymbol{\hat{\ell}})=-{\mu}_{\beta\nu\alpha}^\mathrm{dr}(-\boldsymbol{\hat{\ell}})$, and $\mu_{\alpha\beta\gamma\lambda}^\mathrm{dd}(\boldsymbol{\hat{\ell}})=\mu_{\gamma\lambda\alpha\beta}^\mathrm{dd}(-\boldsymbol{\hat{\ell}})$. The overall minus sign for the relation between the rotational-dipolar components is a consequence of ${\bf T}$ being a pseudovector. Moreover, the remaining components of $\boldsymbol{\zeta}$ follow from the partial inversion of $\boldsymbol{\mu}$, which determine the viscous dissipation of a particle with given $\boldsymbol{\Omega}$ and $\bm{\mathsf{E}}^\infty$. 

From the expression of $\boldsymbol{\mu}^\mathrm{rr}$ in Eq. \eqref{eq:mudd} , we find that $\boldsymbol{\zeta}^\mathrm{rr}$ has the elegant form
\begin{gather}
\boldsymbol{\zeta}^\mathrm{rr}={8\pi\eta_\mathrm{s}a^3}\left[\bm{\mathsf{I}}-\frac{\gamma}{2}(\boldsymbol{\epsilon}\cdot\boldsymbol{\hat{\ell}})
\right].
\end{gather}
Consequently, there is no enhanced rotational dissipation compared to the Stokes case for any direction of $\boldsymbol{\Omega}$ with respect to $\boldsymbol{\hat{\ell}}$, contrasting the translating case. This observation was also noted in a recent preprint using a different argument, see Ref. \cite{Vilfan:2024}. Furthermore, we find
\begin{gather}
\boldsymbol{\zeta}^\mathrm{dr}=4\pi\eta_\mathrm{s}a^3\gamma \left(\bm{\mathsf{p}}^{(0)}+\frac{1}{2}\bm{\mathsf{p}}^{(1)}\right),
\end{gather}
With the given $\boldsymbol{\zeta}^\mathrm{dr}$, $\boldsymbol{\zeta}^\mathrm{rr}$, and the singularity representation of the fluid velocity field Eq. \eqref{eq:velovelo}, we find that a rotating sphere with angular velocity $\boldsymbol{\Omega}$ in the absence of an ambient flow field is described by a rotlet velocity field, ${{\bf v(r)}=-a^3\boldsymbol{\Omega}\times\nabla\left(1/r\right)}$. This is in agreement with the findings in Ref. \cite{Vilfan:2024} obtained via a different route.


\underline{\emph{Effective stress tensor}.} -- 
Next, we consider a suspension of $N$ spherical particles in an OV liquid of volume $V$, with number density $\rho=N/V$. 
 When we neglect hydrodynamic interactions between the spherical particles --valid for sufficiently dilute systems \cite{Beenakker:1984}--, the effective stress tensor can be found via a volume average, see Ref. \cite{Kim}. In our case, we find that the effective stress tensor is $\boldsymbol{\sigma}^\mathrm{eff}=-{p}^\mathrm{eff}\bm{\mathsf{I}}+\boldsymbol{\eta}^\mathrm{eff}:\langle{\nabla{\bf v}}\rangle$, with ${p}^\mathrm{eff}$ a renormalized effective pressure and angular brackets denoting a volume average. Here, the effective viscosity tensor is $\boldsymbol{\eta}_\mathrm{eff}=\boldsymbol{\eta}+\rho\boldsymbol{\mu}^\mathrm{dd}$, which has the mathematical structure
\begin{align}
\boldsymbol{\eta}^\mathrm{eff}(\rho,\gamma)=& [2\eta_\mathrm{s}+\rho\alpha_0(\gamma)]\bm{\mathsf{d}}^{(0)}+[2\eta_\mathrm{s}+\rho\alpha_1(\gamma)]\bm{\mathsf{d}}^{(1)}\nonumber \\
&+[2\eta_\mathrm{s}+\rho\alpha_2(\gamma)]\bm{\mathsf{d}}^{(2)}
+[2\eta_\mathrm{o}+\rho\beta_1(\gamma)]\bm{\mathsf{c}}^{(1)}\nonumber\\
&-[2\eta_\mathrm{o}-\rho\beta_2(\gamma)]\bm{\mathsf{c}}^{(2)} \label{eq:effstress},
\end{align}
with coefficients $\alpha_i(\gamma)$ and $\beta_i(\gamma)$ given by the form of $\boldsymbol{\mu}^\mathrm{dd}$, see Eq. \eqref{eq:mudd}. 
For $\gamma\rightarrow 0$, we retrieve the Einstein result \cite{Einstein:1906, Einstein:1911} $\boldsymbol{\eta}=2\eta_\mathrm{s}[1+(5/2)\varphi]\boldsymbol{\mathcal{I}}$, with $\varphi=(4/3)\pi a^3\rho$ the volume fraction. 

{\color{black} Eq. \eqref{eq:effstress} has the following interpretation. We started with the microscopic structure of a system with self-spinning particles (Fig. \ref{fig:scheme}). On the mesoscopic scale with respect to these particles, we described this system as a fluid governed by hydrodynamic equations with a simplified (but derivable \cite{Markovich:2021}) viscosity tensor characterised by one shear viscosity and one odd viscosity coefficient. Subsequently, we considered a dilute suspension of passive spherical particles in such fluid. On the macroscopic scale where the whole suspension is treated as a continuous fluid \cite{happel}, we end up with an effective hydrodynamic description with three shear viscosities since $\alpha_i(\gamma)\neq 0$ ($i=0,1,2$) and two odd viscosities since $\beta_j(\gamma)\neq 0$ ($j=0,1$), contrasting the simplified viscosity tensor [Eq. \eqref{eq:visctens}] of the fluid on the mesoscopic level.} In the notation of Ref. \cite{Khain:2022}, where a different basis for $\boldsymbol{\eta}$ is used, we generated non-zero shear viscosities $\mu_1$, $\mu_2$, and $\mu_3$, and two odd viscosity coefficients $\eta_1^\mathrm{o}$ and $\eta_2^\mathrm{o}$.
{\color{black} We conclude that the inherent length scale on which hydrodynamics is considered, can thus be utilised in designing suspensions with tailored rheological properties.}

\underline{\emph{Discussion}.} -- {\color{black} Our work provides exact analytical results for interpreting experiments of three-dimensional chiral active fluids with large OV coefficients.} Furthermore, there are important consequences of our results. First, our exact analytical solutions are a starting point for describing hydrodynamic interactions in OV fluids beyond the linear $\gamma$ approximation, where indirect dissipative OV effects will contribute. Second, our results open the doors to describe (many-body) Brownian motion in a system of self-spinning particles. Third, the singularity method that we used in this work can be employed to derive the grand mobility matrix of spherical particles in systems with spatially constant $\boldsymbol{\eta}$. Examples are nematic liquid crystals with a constant director field with strong anisotropy \cite{Stark:2001, Kos:2018} {\color{black} or OV fluids with non-vanishing rotational viscosities, which is relevant for experimental systems \cite{Soni:2019}. In these cases, one  needs to solve the torque balance in addition to the balance of linear momentum. Finally, we discussed the novel notion of a suspension with a chiral active fluid acting as the medium.} Our results on the effective viscosity tensor for the macroscopic hydrodynamics of such suspensions suggest that one can design various complex fluids by mixing self-spinning active fluids with passive particles, which can be important for microfluidic and biological applications.  

\underline{\emph{Acknowledgements}.} --  The authors acknowledge P.~Szymczak, P. Surówka, and F. Pena-Benitez for useful discussions. We are grateful to A. Vilfan for addressing and discussing a mistake in an earlier version concerning the expression for $\boldsymbol{\mu}^\mathrm{rr}$. Furthermore, we acknowledge {Y.~Hosaka} for thoroughly checking our results, which helped us to find the error in our calculations.
\bibliography{literature1} 
\end{document}